\documentclass[twocolumn,prl,aps,footinbib,amsmath,amssymb,superscriptaddress,longbibliography]{revtex4-1}
\usepackage{array}
\usepackage{graphicx}
\usepackage{dcolumn}
\usepackage{color}
\usepackage{bm}
\usepackage{natbib,hyperref}
\usepackage{amsmath,amssymb,amsfonts,bbold}
\usepackage[normalem]{ulem}
\usepackage{setspace}

\newcommand{\beq}{\begin{equation}}
\newcommand{\eeq}{\end{equation}}
\newcommand{\beqa}{\begin{eqnarray}}
\newcommand{\eeqa}{\end{eqnarray}}

\usepackage[dvipsnames]{xcolor}

\graphicspath{{fig/}}
 
\begin{document}
\title{Observation of tunable single-atom Yu-Shiba-Rusinov states}

\author{Artem B. Odobesko}
\email[Corresponding author for experiment: ]{Artem.Odobesko@physik.uni-wuerzburg.de}
\affiliation{Physikalisches Institut, Experimentelle Physik II, 
		Universit\"{a}t W\"{u}rzburg, Am Hubland, 97074 W\"{u}rzburg, Germany}
\author{Domenico Di Sante} 
\email[Corresponding author for theory: ]{Domenico.DiSante@physik.uni-wuerzburg.de}
\affiliation{Institut f{\"u}r Theoretische Physik und Astrophysik, 
	Universit\"{a}t W\"{u}rzburg, Am Hubland, 97074 W\"{u}rzburg, Germany}
\author{Alexander~Kowalski}
\affiliation{Institut f{\"u}r Theoretische Physik und Astrophysik, 
	Universit\"{a}t W\"{u}rzburg, Am Hubland, 97074 W\"{u}rzburg, Germany}
\author{Stefan Wilfert}
\affiliation{Physikalisches Institut, Experimentelle Physik II, 
		Universit\"{a}t W\"{u}rzburg, Am Hubland, 97074 W\"{u}rzburg, Germany}
\author{Felix Friedrich}
\affiliation{Physikalisches Institut, Experimentelle Physik II, 
		Universit\"{a}t W\"{u}rzburg, Am Hubland, 97074 W\"{u}rzburg, Germany}
\author{Ronny Thomale}
\affiliation{Institut f{\"u}r Theoretische Physik und Astrophysik, 
	Universit\"{a}t W\"{u}rzburg, Am Hubland, 97074 W\"{u}rzburg, Germany}
\author{Giorgio Sangiovanni}
\affiliation{Institut f{\"u}r Theoretische Physik und Astrophysik, 
	Universit\"{a}t W\"{u}rzburg, Am Hubland, 97074 W\"{u}rzburg, Germany}
\author{Matthias Bode} 
	\affiliation{Physikalisches Institut, Experimentelle Physik II, 
	Universit\"{a}t W\"{u}rzburg, Am Hubland, 97074 W\"{u}rzburg, Germany}	
	\affiliation{Wilhelm Conrad R{\"o}ntgen-Center for Complex Material Systems (RCCM), 
	Universit\"{a}t W\"{u}rzburg, Am Hubland, 97074 W\"{u}rzburg, Germany}   

\date{\today}

\begin{abstract} 
The coupling of a spin to an underlying substrate is the basis for a plethora of phenomena. 
In the case of a metallic substrate, Kondo screening of the adatom magnetic moment can occur. 
As the substrate turns superconducting, an intriguing situation emerges where the pair breaking 
due to the adatom spins leads to Yu-Shiba-Rusinov bound states, but also intertwines 
with Kondo phenomena. 
Through scanning tunneling spectroscopy, we analyze the interdependence of Kondo screening and superconductivity.
Our data obtained on single Fe adatoms on Nb(110) show that the coupling and the resulting YSR states 
are strongly adsorption site-dependent and reveal a quantum phase transition 
at a Kondo temperature comparable to the superconducting gap.
The experimental signatures are rationalized by combined density functional theory
and continuous-time quantum Monte-Carlo calculations to rigorously treat magnetic and hybridization effects on equal footing. 
 
\end{abstract}
\maketitle

\emph{Introduction}---Chains of magnetic atoms on the surface of a highly spin-orbit coupled superconductor
were the basis for the 
discovery of Majorana zero modes (MZM) 
which have been predicted \cite{Nadj-Perge2013,Pientka2013,Klinovaja2013,Sarma2014} 
and reported \cite{Yazdani2014,Kim2018,Kamlapure2018} to exist at the ends 
of one-dimensional (1D)  spin systems. 
Constant efforts are undertaken for improving the properties of such hybrid systems. 
One 
crucial aspect is to precisely control the chain structure and length, 
which has been achieved by atomic manipulation on refractory substrates \cite{Kim2018}.
Another promising though hitherto not largely explored direction 
is to increase the critical temperature $T_{\rm c}$ of the superconductor.
With its high $T_{\rm c} = 9.26$~K and its shallow surface atomic potential, 
the 
Nb(110) surface offers ideal properties for achieving both, 
a high spectral resolution and atomic manipulation, thereby representing a
promising alternative to both Pb \cite{Yazdani2014} and Re \cite{Kim2018}.

Another important parameter is how the magnetic adatoms couple to the substrate. 
The resulting local moment and its screening by the substrate's conduction electrons
is described by the hybridization function. 
In the normal-metallic phase ($T > T_{\rm c}$) the screening can lead to a drastic reduction 
of the impurity spin or its complete screening  by a Kondo singlet cloud \cite{Kondo1964}. 
Below $T_{\rm c}$, the Kondo effect intertwines with the formation of localized bound resonances, 
known as Yu-Shiba-Rusinov (YSR) states~\cite{Yu,Shiba,Rusinov}.
These YSR states produce trivial low-energy features which may be hardly distinguishable 
from topologically protected MZM \cite{Ruby2015,Cornils2017,Yazdani2017}. 

\begin{figure*}[t]  
		\includegraphics[width=1\textwidth]{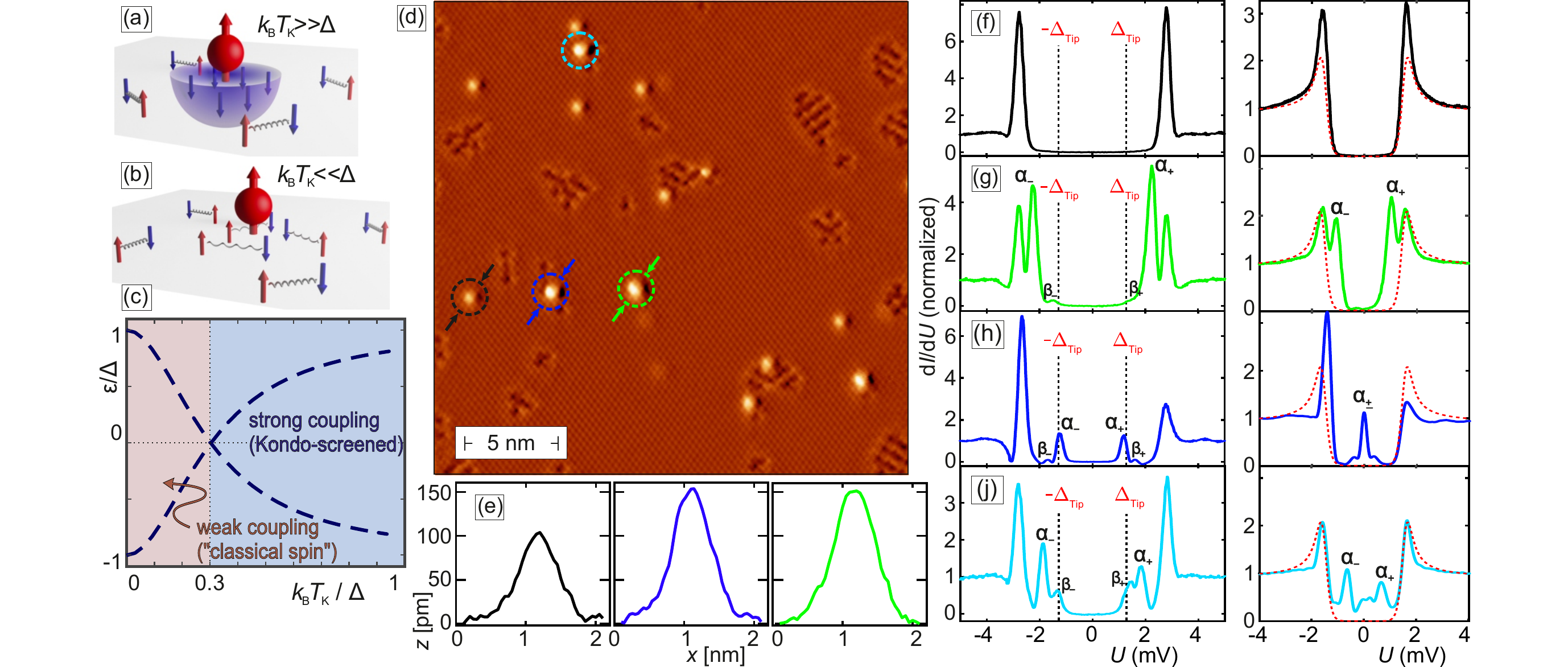}
		\caption{(a) Sketch of YSR states in dependence on the effective exchange
			interaction between a magnetic impurity with $S = 1/2$ and a superconducting substrate 
			for strong coupling ($k_{\rm B} T_{\rm K}  \gg \Delta$):   
			the impurity spin is fully screened with a ground state spin $S \approx 0$. 
			(b) Same as (a) but for weak coupling ($k_{\rm B} T_{\rm K}  \ll \Delta$):   
			the local spin is only partially screened resulting in an $S \approx 1/2$ many-body ground state.
			(c) Treating the spin as a classical moment, the position of the YSR states 
			inside the superconducting gap follow the dashed lines.
			(d) STM image of Fe adatoms on Nb(110) 
			(stabilization parameters: $U = -0.6$\,V, $I = 0.4$\,nA). 
				(e) Line profiles measured between the arrows in (a) 
			for Fe/Nb(110) at a H$_4$ site (black), 
			on top or close to NbO$_x$ (blue), and Fe/Nb(110) in a bridge site (green). 
			The two columns on the right show as measured and deconvoluted spectra 
			obtained with a superconducting tip on faint and bright Fe atoms on the clean Nb(110) surface (f,g); 
			and typical spectra measured on Fe atoms on NbO$_x$ (h,j). 
			Line colorcode matches atom color markers on STM image (d). The Nb spectrum is shown as a hatched red line.
			Stabilization parameters: $U_{\textrm{set}} = -7$\,mV, $I_{\textrm{set}} = 0.4$\,nA.}
		\label{fig:STM_STS} 
\end{figure*}

Two limiting situations 
are sketched in Fig.\,\ref{fig:STM_STS}. 
In the strong coupling regime with $k_{\rm B} T_{\rm K}  \gg \Delta$, where $k_{\rm B}$ is the Boltzmann constant, 
$T_{\rm K}$ the Kondo temperature, and $\Delta$ the superconducting gap, 
magnetic scattering results in pair-breaking and leads to a Kondo resonance 
with a characteristic energy scale $k_{\rm B} T_{\rm K}$ \cite{Matsuura1977}. 
In the simplified picture of an $S = 1/2$ impurity this corresponds to an $S = 0$ 
many-body ground state, depicted in Fig.\,\ref{fig:STM_STS}(a). 
Conversely, at very weak coupling ($k_{\rm B} T_{\rm K}  \ll \Delta$) the pair-breaking effect 
almost vanishes [Fig.\,\ref{fig:STM_STS}(b)] and YSR bound states appear 
close to the edge of the superconducting gap. 
The intermediate situation ($k_{\rm B} T_{\rm K}  \approx \Delta$) 
is determined by an involved interplay between the two effects.
Theory predicts a transition between the two limiting cases at $k_{\rm B} T_{\rm K}  \approx 0.3 \Delta$, 
where the two YSR resonances cross \cite{Satori1992, Sakai1993, zitkoPRB83}, Fig.\,\ref{fig:STM_STS}(c). 
In this picture, which has so far only been verified for molecular layers \cite{Franke2011,Hatter2017}, 
the Kondo effect can be considered as pair-breaking 
and a quantity for the in-gap position of YSR resonances.

Here we present scanning tunneling spectroscopy (STS) data for single Fe adatoms 
on atomically clean Nb(110) and oxygen-reconstructed surfaces (NbO$_x$)
of the elemental superconductor Nb. 
We find that the in-gap position of YSR states critically depends on the adsorption site,  
with a much stronger coupling of Fe to NbO$_x$ than to clean Nb(110). 
Combined first-principles density functional theory (DFT) and continuous-time quantum Monte Carlo (CT-QMC) calculations 
show that a subtle reduction of the Fe--NbO$_x$ binding distance
results in an enhanced hybridization of atomic orbitals, 
sizable enough to overcompensate the oxygen-induced reduction 
of the Fe magnetic moment, eventually resulting in an enhanced $T_{\rm K}$.
The systematic understanding of trivial screening effects of magnetic moments 
that hybridize with normal-metallic and superconducting surfaces represents 
an important corner stone toward engineered topological devices. 

\emph{Experimental  and theoretical methods}---A clean Nb(110) surface with less than 
10\% NbO$_{x}$ was prepared by heating up to $T = 2410^\circ$C \cite{Odobesko2019}. 
Scanning tunneling microscopy (STM) measurements were performed 
at $T_{\rm min} = 1.17$\,K. 
Fe atoms were evaporated \textit{in-situ} onto cold Nb surface at $T = 4.2$\,K. 
For increased energy resolution, the W tips  were gently poked into the Nb sample to create a superconducting tip apex,  
resulting in a shift of all spectral features by $\pm \Delta_{\rm tip}/e$.
DFT calculations were performed by using the VASP simulation package~\cite{PhysRevB.54.11169},
while the solution of the Anderson impurity model (AIM) was achieved
by the CT-QMC method within the w2dynamics code \cite{w2dynamics}. 
We have used a uniform Kanamori interaction parameterized 
by $U = 3.85$\,eV and $J^{\text{Hund}} = 0.72$\,eV \cite{Hausoel2017}. More details are given in Ref.~\cite{SupplMat}.


\emph{Results}---One major obstacle toward a systematic investigation of a direct correlation 
between the Kondo effect and YSR states is the difficulty in tuning the adatom--substrate coupling.   
A single magnetic Fe atoms adsorbed on top of clean and oxygen-reconstructed Nb(110)\cite{Odobesko2019} 
offer a wide range of substrate-tunability which allows for a critical evaluation of the above-mentioned scenarios.  
Fig.\,\ref{fig:STM_STS}(d) shows an atomic resolution STM image 
of a typical Nb(110) surface\cite{Odobesko2019} after low-temperature deposition of Fe. 
Since the surface is not perfectly clean but contains a few oxygen-reconstructed patches (dark stripes),
Fe adatoms in different adsorption sites can be found.  
Line sections drawn across three Fe atoms marked by black, green, or blue arrows 
in Fig.\,\ref{fig:STM_STS}(d) are displayed in Fig.\,\ref{fig:STM_STS}(e). 
Obviously, the respective corrugation heights are significantly different.  
The 'black' atom adsorbed in a four-fold hollow ($H_4$) site 
of clean Nb(110) exhibits a height of $\approx 100$\,pm.  
In contrast, the 'blue' atom on NbO${_x}$ and the 'green' atom on clean Nb(110) at an unconventional binding site
both are $\approx 150$\,pm high and therefore appear brighter in Fig.\,\ref{fig:STM_STS}(d).

\begin{figure*} [t]
		\includegraphics[width=1\textwidth]{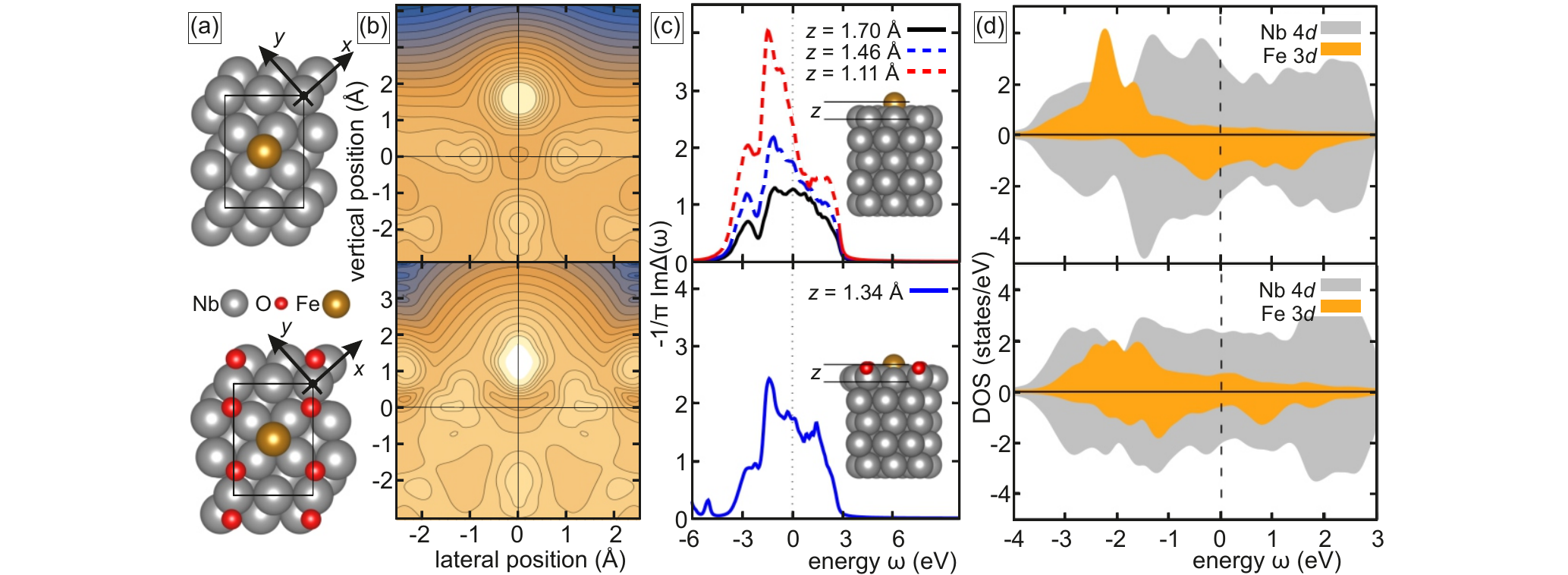}
		 \caption{{Theoretical investigation of Fe atoms 
			on clean (top) and oxygen-reconstructed (bottom) Nb(110).} 
			(a) Top view unit cells used for {\em ab initio} calculations. 
			(b) Charge density profiles.  
			The higher charge density over Fe/NbO$_x$ translates into its larger apparent height 
			in constant-current STM images [cf.\,Fig.\ref{fig:STM_STS}], despite its lower binding distance.
			(c) Imaginary part of the hybridization function. The three curves in the top panel 
			refer to three Fe adatom adsorption heights on the clean Nb(110) surface. 
			The equilibrium geometry ($z = 1.70$\AA) is shown as a solid black curve.
			(d) DOS projected on the surface Nb (grey) and Fe (orange) $d$-states. 
			Positive and negative values correspond to majority and minority spin, respectively.}
		\label{fig:theory}  
\end{figure*} 

These different adsorption sites substantially impact the interaction with the substrate, 
as revealed by tunneling spectra measured above the respective Fe atoms. 
The spectrum measured on the 'black' Fe atom qualitatively resembles the spectrum 
of pristine Nb (red hatched line) with enhanced 
coherence peaks, Fig.\,\ref{fig:STM_STS}(f).  
This observation corresponds to the weak coupling limit described above, 
suggesting an extremely small exchange coupling with the superconductor.  
In contrast, all other types of Fe adatoms exhibit additional in-gap peaks. 
For example, the rare case with 'green' atom on clean Nb(110) 
shows fingerprints of YSR states close to the gap edge strongly pronounced $\alpha_{\pm}$
and much weaker $\beta_{\pm}$ peaks, Fig.\,\ref{fig:STM_STS}(g).  
The electron--hole ($e^-$--$h^+$) asymmetry of the spectral density of the YSR resonances
originates from the local crystal field acting on the adatom from the Coulomb multiplet 
of the Fe $d$-shell\cite{Balatsky1997,FlattePRB1997,Franke2011,Cornils2017}.  
Typical tunneling spectra of Fe atoms sitting on different sites 
within the NbO$_x$ unit cell are presented in Fig.\,\ref{fig:STM_STS}(h),(j).  
As compared to Fig.\,\ref{fig:STM_STS}(g), in Fig.\,\ref{fig:STM_STS}(h) 
we find the $\alpha_{\pm}$ peaks further inside the superconducting gap. 
In contrast, the in-gap position of the weaker $\beta_{\pm}$ peaks remains almost unchanged.
The deconvoluted data in Fig.\,\ref{fig:STM_STS}(h) reveal that the $\alpha_{\pm}$ peaks 
coincide at the Fermi level ($E_{\rm F}$), a situation hardly distinguishable from a MZM.
In Fig.\,\ref{fig:STM_STS}(j) the spectral density of the $\alpha_{\pm}$ peaks 
appears with an inverted $e^-$--$h^+$ asymmetry, 
indicating a transition toward the strong coupling limit. 


The position of YSR states inside the superconducting gap 
is given by $\varepsilon = \Delta \left[ (1 - a^2) /  (1 + a^2) \right]$, 
where $a = J m \pi \rho_s$ depends on the exchange interaction $J$, the magnetic moment $m$, 
and the normal metallic density of states (DOS) $\rho_s$ at $E_{\rm F}$\cite{Satori1992,Sakai1993,zitkoPRB83,Hatter2017}.
The presence of non-metallic adsorbates affect the moment of magnetic materials \cite{Getzlaff1999}
and in many cases causes a significant reduction \cite{PhysRevLett.71.2122,PhysRevLett.76.2802}. 
Along this line of reasoning the observation of a stronger coupling on NbO$_x$ is at first surprising.  
To clarify this we performed the combined DFT and CT-QMC calculations 
which allow us go beyond the classical description and thereby also access the strong coupling regime. 
As we will show below this gives full account of the many-body properties of the interacting impurity 
on the Nb substrate even without explicitly considering its superconducting phase.

The first step is to investigate single Fe atoms in their characteristic adsorption sites by DFT.  
The respective unit cells 
are presented in Fig.\,\ref{fig:theory}(a).
We find equilibrium distances of 170\,pm for Fe/Nb(110) and 134\,pm for Fe/NbO$_x$. 
As confirmed by the calculated charge density profiles presented in Fig.\,\ref{fig:theory}(b), 
this leads to a higher charge density for the Fe atom on NbO$_x$ than on clean Nb(110), 
which---in the constant-current mode of the STM---translates into a larger apparent height, 
as observed experimentally [cf.\ Fig.\,\ref{fig:STM_STS}(e)].   
This variation of binding distance crucially impacts the adatom--substrate coupling. 
Close to $E_{\rm F}$ we observe a changing form and value 
of the hybridization function $\Delta(\omega)$ [Fig.\,\ref{fig:theory}(c)], 
which describes the probability amplitude of an electron to hop from the impurity to the substrate and back 
and defines a multi-orbital Anderson impurity model. 
As evidenced by the DFT majority and minority spin DOS shown in Fig.\,\ref{fig:theory}(d), 
the moment of Fe/NbO$_x$ ($m^{*} = 0.9\mu_{\rm B}$) amounts less than half 
of the corresponding value for Fe/Nb(110) in the $H_4$ site ($m = 2.2\mu_{\rm B}$). 
At the same time, the hybridization function steeply increases [cf.\ Fig.\,\ref{fig:theory}(c)], 
leading to a larger exchange interaction $J$ for Fe/NbO$_x$.  
As will be shown this overcompensates the reduction of the impurity moment, 
and bring the system into the Kondo-screened regime.

\begin{figure}[t] 
		\includegraphics[width=\columnwidth]{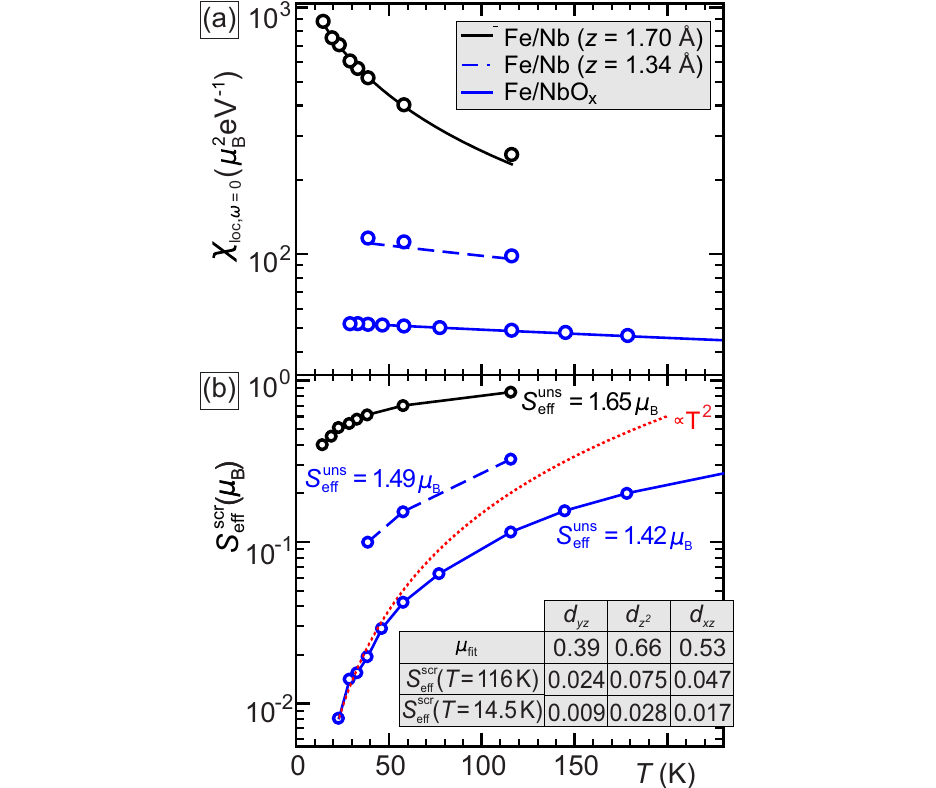}
		\caption{
     			Temperature-dependent static local spin susceptibility (a) and effective impurity spin (b) 
			of Fe/Nb(110) (distances $z = 1.70${\AA} and $z = 1.34${\AA}) and for Fe on NbO$_x$. 
			Black line: Curie-Weiss fit with Wilson form $\mu_\text{fit}^2/3(T + 2 T_{\rm K})$.   
			The unscreened effective spin ($S_\text{eff}^\text{uns}$) is $T$ independent, 
			representing the instantaneous paramagnetic moment. 
			When lowering $T$, screening via the substrate becomes increasingly effective, 
			as shown by the $S_\text{eff}^\text{scr}(T)$ behavior 
			obtained from the spin--spin correlation function at long (imaginary) times. 	 
			Inset: Values of $\mu_\text{fit}$ and the screened effective spin $S_\text{eff}^\text{scr}$ 
			(both in units of $\mu_{\rm B}$) for three orbitals of the $d$ shell, calculated at $z = 1.70$\,{\AA}.}
		\label{fig:theory_spin} 
\end{figure}

The $T$ dependences of the impurity spin susceptibility $\chi_{\text{loc}, \omega\!=\!0}$ 
and the effective screened spin $S_\text{eff}^\text{scr}$ of the Fe impurity 
as calculated with CT-QMC are shown in Fig.\,\ref{fig:theory_spin}. 
For a free local moment, $\chi_{\text{loc}, \omega\!=\!0}(T)$ is of Curie-Weiss type ($\propto 1/T$) 
and changes to a weakly $T$-dependent Pauli-like behavior for moments screened by conduction electrons.
Although the Fe impurity will always be fully screened at $T = 0$,
the way this screening takes place when lowering $T$ can vary significantly. 
For the $T$ range investigated in our calculations, Fe on clean Nb displays a Curie-Weiss susceptibility
and the corresponding effective spin is still far from being fully screened. 
On the contrary, $\chi_{\text{loc}, \omega\!=\!0}(T)$ for Fe/NbO$_x$ 
exhibits a pronounced Pauli-like behavior and $S_\text{eff}^\text{scr}(T)$ is significantly reduced.

As shown in Fig.\,\ref{fig:theory_spin}, our analysis of the crossover between these two physical regimes 
identifies the Fe--surface distance $z$ as the driving mechanism. 
This is confirmed by simulating Fe on Nb(110) at an artificial height. 
Selecting a value comparable to Fe/NbO$_x$---see blue dashed line in panel (a)---we observe 
a qualitatively similar Pauli spin response and a low $S_\text{eff}^\text{scr}$. 
$T$-dependent $S_\text{eff}^\text{scr}$ data also allow to extract hints on the respective $T_{\rm K}$. 
In Fe/NbO$_x$ an inflection point is visible at around 40\,K. 
Below this temperature $S_\text{eff}^\text{scr}$ decays Fermi liquid-like, 
i.e., $\propto T^2$ [red-dotted line in Fig.\,\ref{fig:theory_spin}(b)]. 
For Fe/Nb(110) no such inflection point is reached down to the lowest $T$, 
confirming that Fe/Nb(110) and Fe/NbO$_x$ belong to different Kondo-coupling regimes.

Additionally, for Fe on clean Nb(110) at $z = 1.70${\AA}, 
given the pronounced Curie-Weiss $\chi_{\text{loc}, \omega\!=\!0}(T)$, 
we can fit the CT-QMC data to $\mu_\text{fit}^2/3(T + 2 T_\text{K})$, 
a formula valid for an intermediate temperature range \cite{RevModPhys.47.773}. 
The resulting susceptibility which includes contributions from the whole $d$ shell 
is shown as a black solid line in Fig.\,\ref{fig:theory_spin}(a). 
Analyzing the individual, orbital-specific contributions to $\chi_{\text{loc}, \omega\!=\!0}$ 
we see that $\mu_\text{fit}$ for $d_{z^2}$ is larger than for $d_{xz,yz}$, 
see table inset to Fig.\,\ref{fig:theory_spin}(b). 
We infer that the former hosts the most correlated electrons, 
also confirmed by the orbital-resolved values $S_\text{eff}^\text{scr}$%
~\footnote{In this comparison we did not consider the two purely planar orbitals $d_{xy}$ and $d_{x^2-y^2}$ 
since they are less relevant for tunneling into the tip.}.
The fit yields a $T_\text{K}$ value consistent with an $S_\text{eff}^\text{scr}(T)$ inflection point 
of the order of 10\,K, i.e., four times smaller than for Fe/NbO$_x$.  

A well-known characteristic of correlated electrons is their strong response 
to small changes of parameters \cite{Imada1998, Limelette2003, Kuegel2018}. 
Given its out-of-plane orientation, we expect that $d_{z^2}$ states is most susceptible 
to geometrical changes such as a variations of the adatom--substrate distance $z$. 
Therefore, the highly responsive $\alpha_{\pm}$ peaks probably originate 
from processes involving the $d_{z^2}$ orbital.
For the $\beta_{\pm}$ peaks we have less stringent evidence, 
but most plausibly they are related to $d_{xz,yz}$ orbitals %
\footnote{Let us note at this point that, while the previous orbital-resolved analysis is informative, 
one has to bear in mind that the diagonal contributions to $\chi_{\text{loc}, \omega\!=\!0}$ 
and $S_\text{eff}^\text{scr}$ constitute only a small fraction of the respective full values 
[compare table in Fig.\,\ref{fig:theory_spin}(b) with the CT-QMC curves shown in the figure)]. 
This means that the Kondo effect in these adatoms ultimately involves five entangled 
(and partially-filled) $d$-orbitals and its picture as five independent Kondo channels 
is to be taken \emph{cum grano salis}.}. 

\begin{figure}[t] 
		\includegraphics[width=\columnwidth]{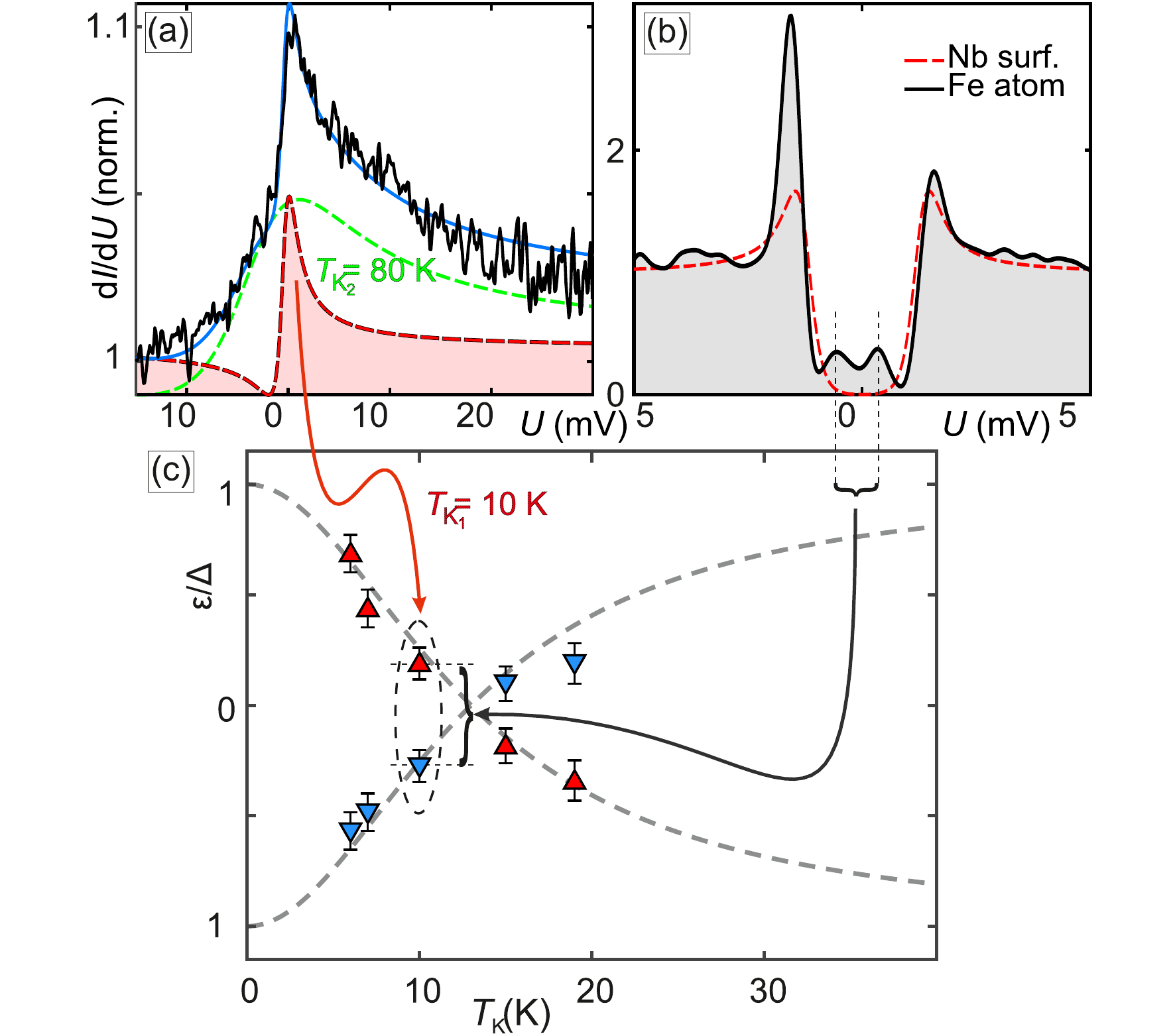}
	\caption{
		Kondo resonance (a) and YSR bound states (b) measured on the same Fe atom 
		at $T=1.5$~K with and without external magnetic field $\mu_0 H=0.6$\,T, respectively. 
		The Kondo resonance is fitted with two Fano curves, with low $T_{\rm K_1}=10$\,K (red dashed line) 
		and high $T_{\rm K_2}=80$\,K (green). Blue: Sum of both Fano curves. 
		(c) $T_{\rm K}$-de\-pen\-dent positions of the YSR peaks measured for five individual Fe atoms. 
		The transition occurs at $T_{\rm K_1}=13$\,K. \label{fig:YSR_TKondo} } 
\end{figure} 
Remarkably, we can draw quantitative conclusions on YSR states for the \emph{superconducting} phase
even though $T_\text{K}$ is obtained form many-body correlations in the \emph{normal} state \cite{zitkoPRB83}. 
To verify this claim of a direct relation between the substrate-tuned Kondo temperature and YSR states,
we performed two-stage STS measurements on the same set of five individual Fe atoms.  
At the first stage we investigated the Kondo effect on normal-metallic Nb(110).   
To suppress the superconducting state the sample was exposed 
to an out-of-plane magnetic field $\mu_0 H = 0.6$\,T\,$> \mu_0 H_{\rm c_2}$. 
A spectrum with the characteristic asymmetric Kondo peak is presented in Fig.\,\ref{fig:YSR_TKondo}(a).  
In line with the theoretical discussion above it can be fitted by two Fano curves, 
accounting for the interaction of the $d_{z^2}$ and $d_{xz,yz}$ orbitals, respectively.  
Whereas the feature with the broader peak exhibits a rather high $T_{\rm K_2} = (75 \pm 10)$\,K 
which is---within our error bar---independent of the particular Fe atom adsorption site, 
the second, more narrow feature has a much lower Kondo temperature ($T_{\rm K_1}$) 
which varies significantly between the Fe adatoms. 
For example, $T_{\rm K_1} = 10$\,K for the particular Fe atom 
analyzed in Fig.\,\ref{fig:YSR_TKondo}(a).

At the second stage the magnetic field was turned off and
the same Fe adatoms were investigated on superconducting Nb(110). 
The deconvoluted tunneling spectrum of the same Fe adatom 
analyzed in Fig.\,\ref{fig:YSR_TKondo}(a) is presented in Fig.\,\ref{fig:YSR_TKondo}(b). 
We recognize a pronounced superconducting gap and YSR in-gap peaks.  
In Fig.\,\ref{fig:YSR_TKondo}(c) we summarize the data obtained 
from five Fe adatoms by plotting their individual YSR peak positions 
versus the respective lowest Kondo temperature $T_{\rm K_1}$ \cite{SupplMat}. 
We find a good qualitative agreement with theoretical predictions\,\cite{Satori1992,Sakai1993,zitkoPRB83}, 
as the transition point occurs at a critical Kondo temperature $T_{\rm K_c} \approx 0.7\Delta$. 
This value is similar but slightly below the pairing parameter $\Delta$ reported for moir{\'e} patterns 
of Mn-phthalocyanine molecules on Pb(111) \cite{Franke2011, Bauer2013}. 
Even though alternative definitions of $T_{\rm K}$ modify the position of the transition \cite{Hewson1993}, 
the crossover always remains below the pairing parameter $\Delta$, thereby suggesting a universal, 
largely template-independent quantum phase transition 
at a Kondo energy scale corresponding to the superconducting gap.

Beyond the detailed physical understanding the results presented here may guide future efforts in creating
atomic scale assemblies which hold  MZM. A necessary condition is a strong coupling between the superconductor
and the magnetic impurities,  which is facilitated for Fe adatoms on Nb(110) by introducing an oxygen in between.
Wherein $T_{\rm c}$ remains essentially unaffected such that the advantages of Nb can still be fully exploited.


\emph{Acknowledgments}---Work was supported by the Deutsche Forschungsgemeinschaft 
(DFG, German Research Foundation) -- Project-ID 258499086 -- SFB 1170 (A02/B06/C07) 
and through the W\"urzburg-Dresden Cluster of Excellence on Complexity and Topology
in Quantum Matter -- \textit{ct.qmat} Project-ID 39085490 - EXC 2147.
The authors acknowledge R. \u{Z}itko for useful discussions, and the Gauss Centre for
Supercomputing e.V. for providing computing time on the GCS Supercomputer
SuperMUC at Leibniz Supercomputing Centre.

\bibliography{Bibliography_YSR_07}

\end{document}